\documentclass[11pt]{article}
\usepackage{amsfonts}

\newcommand{\ka}{{\bf A}}
\newcommand{\kk}{{\bf K}}
\newcommand{\kc}{{\bf K_C}}
\newcommand{\kd}{{\bf K_D}}
\newcommand{\bw}{{\it{Proof.}}\hspace{3mm}}
\newcommand{\bt}{\begin{tabular}}
\newcommand{\et}{\end{tabular}}
\newcommand{\ba}{\begin{array}}
\newcommand{\ea}{\end{array}}
\newcommand{\ak}{a^{(k)}}
\newcommand{\de}{}

\def\rr{{\mathbb R}}
\def\qq{{\mathbb Q}}
\def\zz{{\mathbb Z}}
\def\nn{{\mathbb N}}
\def\ff{{\mathbb F}}
\def\ser{{\mathbb S}}
\def\fqo{\ff_q^\infty}
\def\f2o{\ff_2^\infty}
\def\mod{{\mbox {\rm\ mod\ }}}

\def\ba{\begin{array}}
\def\ea{\end{array}}
\newcommand{\y}{\hspace{1mm}}

\begin{document}

\begin{centering}

{\large{\bf 
Continued Fraction Expansion as Isometry\\ 
The Law of the Iterated Logarithm for\\ 
Linear, Jump,  and 2--Adic  Complexity\\
}}
\vspace*{ 8 mm}

Michael Vielhaber, {\it Member, IEEE}
\footnote{
Supported by Project FONDECYT 2001, No. 1010533 of
CONICYT, Chile
}

\vspace*{ 8 mm}

Instituto de Matem\'aticas\\
Universidad Austral de Chile\\
Casilla 567,
Valdivia\\
{\tt uach@gmx.net}

\end{centering}

\vspace*{ 1 cm}

{\bf{\it\bf Abstract} ---
In the cryptanalysis of stream ciphers and pseudorandom sequences, the notions
of linear, jump, and 2--adic complexity arise naturally to measure the
(non)randomness of a given string.
We define an isometry $\kk$ on $\ff_q^\infty$ that is the precise equivalent
to Euclid's algorithm over the reals to calculate the continued fraction 
expansion of a formal power series. 
The continued fraction expansion  allows to deduce the linear and jump
complexity profiles of the input sequence.
Since $\kk$ is an isometry, the resulting $\ff_q^\infty$--sequence is 
i.i.d.~for
i.i.d.~input. Hence the linear and jump complexity profiles may be modelled via
Bernoulli experiments (for  $\ff_2$: coin tossing), and we can apply the very
precise bounds as collected by {R\'ev\'esz}, among others the Law of the
Iterated Logarithm.

The second topic is the 2--adic span and complexity, as defined by Goresky and
Klapper. We derive again an isometry, this time on the dyadic integers $\zz_2$
which induces an isometry $\ka$ on ${\ff_2}^\infty$. The corresponding jump
complexity behaves on average exactly like coin tossing.

\\\\\\
}

\centerline{\sc{ I. Introduction}}

\centerline{\hspace*{1 mm}}

For some prime power $q$, let $\ff_q$ be the finite field with $q$ elements
\cite{LN}. We consider the set $\fqo$ of infinite sequences over $\ff_q$ as
our starting point. To assess the randomness of such $\fqo$--sequences
one computes their linear and jump complexity profiles.
These profiles as well as the 2--adic complexity should behave well in the 
sense that no large jumps occur and on average, for linear or $2$--adic
complexity, resp.,  $q-1$ out of every $2q$ resp.~$q$  
symbols should lead to a jump.

We shall occasionally employ the set of finite words over $\ff_q$ which we
denote as $\ff_q^*$ (we do not use the multiplicative group of $\ff_q$,
so no confusion should  arise).
In particular, the empty word $\varepsilon$ is in $\ff_q^*$.

Dealing with linear and jump complexity, a sequence
$a=(a_1,a_2,\dots) \in\ff_q^\infty$ is considered as coefficient sequence
of its generating function, the formal power series

$\begin{array}{lcccc}
&G\colon& \fqo&\to& \ser_q\\
&&(a_i)&\mapsto&\sum_{i=1}^\infty a_ix^{-i}
\end{array}$

Here $\ser_q =\{f\in\ff_q[[x^{-1}]] \colon f = \sum_{i=1}^\infty a_ix^{-i}, 
a_i\in\ff_q\}$ is a subring (without unity) of the ring $\ff_q[[x^{-1}]]$
of formal power series. The $f\in\ser_q$ have negative degree.

Furthermore, we use the polynomial ring $\ff_q[x]$ and its field of fractions
$\ff_q(x)$.
Then, $\ser_q\cap \ff_q(x)$ denotes the image of the ultimately periodic sequences in $\fqo$
under $G$.

We define the {\it leading coefficient} $lc(f)$ of a {\it formal power series}
as
$$lc(0)=0 \ {\rm\  and\ }\ lc(\sum_{k=i}^\infty a_kx^{-k})=a_i\in\ff_q
\backslash\{0\},$$
and the {\it degree} as
$$|{\de}0|{\de} = -\infty\ {\rm\  and\ }\ |{\de}\sum_{k=i}^\infty a_kx^{-k}|{\de}=-i
{\rm\  (with\ }a_i\neq 0).$$
The analogue for polynomials is the usual degree  
$|{\de}\sum_{k=0}^d a_kx^{k}|{\de} := d$,
where the leading coefficient $lc(\sum_{k=0}^d a_kx^{k}) := a_d$ again is
assumed nonzero.
The degree fulfills the ultrametric inequality
$|f\pm g| \leq \max\{|f|, |g|\}$  and equality
$|f| \neq |g| \Rightarrow |f\pm g| = \max\{|f|, |g|\}$ 
(we shall not use the associated norm).

A function $f\colon \fqo \to \fqo$ is then called an {\it isometry}, if it 
preserves distance that is for all $a,b\in\fqo\colon |{\de}a-b|{\de} = |{\de}f(a)-f(b)|{\de}$ (coordinate--wise subtraction in $\ff_q$).

The outline of the paper is as follows:

The next section introduces an isometry $\kk$ on $\ff_q^\infty$,  where
$\kk(a)$ describes the partial denominators of the continued fraction
expansion of $G(a)$.  As isometry, \kk{} is information preserving and
we shall see, how each symbol of $\kk(a)$ describes precisely what can be 
said at that moment about the partial denominators.

Section III makes the connection to linear and jump complexity and gives
the wellknown Euclid--Lagrange--Berlekamp--Massey--Dornstetter algorithm
in a form that delivers precisely $\kk(a)$ as discrepancy sequence.

The section IV is a compilation of consequences of \kk{} being an isometry.
Some of these results are already known but the proofs using $\kk$ are shorter.

The main point of the paper is stated in section V. Since $\kk$ is an isometry,
an i.i.d.~sample $a\in\fqo$ leads to an i.i.d.~resulting $\kk(a)$,
and thus on average, $\kk(a)$ can be described by a
Bernoulli experiment (in the binary case $q=2$ just coin tossing)
and we may apply the sharp known bounds (L\'evy classes) for coin
tossing to linear and jump complexity.

Section VI changes the focus to the 2--adic complexity as defined by Klapper
and Goresky. Again, there is an isometry on $\fqo$ associated to 2--adic
complexity and the induced jump complexity turns out to behave
on average just like the ``linear'' jump complexity.

We state an open problem and finish with the conclusion.
\\\\\\
\centerline{\sc II. The Isometry \kk}
\\\\
In this section we will obtain, for a sequence $a\in\fqo$, its generating 
function, the expansion of the latter into a continued fraction, and finally 
an encoding of the partial denominators of the continued fraction. 
The whole process will turn out to define an isometry on $\fqo$, 
which we call $\kk$ (``Kettenbruch'').
\\\\
{\it A.\  Continued Fractions of Formal Power Series}

We start with the continued  fraction expansion of a formal power series
$G(a)\in\ser\backslash\{0\}$.
For $\xi\in\ff_q[[x^{-1}]]$ (not only in $\ser$),
$\xi = \sum_{k=n}^\infty a_kx^{-k}$, we define the integral part as
$\lfloor\xi\rfloor := \sum_{k=n}^0 a_kx^{-k}$. $\lfloor\xi\rfloor$ is zero for
$n\geq 1$ and a polynomial of degree $-n$ for $n\leq 0$.

Given $\xi_0 := G(a)\in\ser\backslash \{0\}$, 
we now proceed iteratively to obtain
$A_i(x) = \lfloor\xi_i\rfloor\in\ff_q[x]$ and
$\xi_{i+1} := (\xi_i - A_i(x))^{-1}\in\ser$. We discard
$A_0(x)$ which is always zero, and obtain a sequence $(A_1(x),A_2(x),\dots)$
of polynomials with positive degree which are  called the 
{\it partial denominators} of the continued fraction expansion.
This sequence is finite, if $G(a) \in\ff_q(x)$ (that
is, if $a$ is ultimately periodic), it is infinite otherwise.
More on continued fractions can be found in Perron \cite{Per} over the reals, 
and in {de Mathan} \cite[ch.~IV]{Mat} or {Artin} 
\cite[\S{}12~f.]{Art} for formal power series.

We can now write the formal power series $G(a)$ as
$$G(a)= \sum_{i=1}^\infty a_i x^{-i} =
\frac{1}{A_1(x)
+ \frac{1}{A_2(x)
+ \frac{1}{A_3(x)+\dots}}}
=:
\frac{\hfill 1\hfill|}{|\hfill A_1(x)\hfill}
+ \frac{\hfill 1\hfill|}{|\hfill A_2(x)\hfill}
+ \frac{\hfill 1\hfill|}{|\hfill A_3(x)\hfill} +\dots$$

More formally, we state this map from $\ser\backslash \{0\}$ to
sequences of polynomials with positive degree in 
$(\ff_q[x]\backslash \ff_q)^* \cup (\ff_q[x]\backslash \ff_q)^\infty$ 
as operator  $\cal K$:

$\begin{array}{lcccc}
&{\cal K}\colon&(\ser_q\cap\ff_q(x))\backslash\{0\} &\to& (\ff_q[x]\backslash \ff_q)^* \\
&&\sum_{i=1}^\infty a_ix^{-i} =\frac{\hfill 1\hfill|}{|\hfill A_1(x)\hfill}
+ \frac{\hfill 1\hfill|}{|\hfill A_2(x)\hfill}
+ \dots
+ \frac{\hfill 1\hfill|}{|\hfill A_k(x)\hfill} &\mapsto& (A_i(x))_{i=1}^k\\
&{\cal K}\colon& \ser \backslash \ff_q(x) &\to& (\ff_q[x]\backslash 
\ff_q)^\infty\\
&&\sum_{i=1}^\infty a_ix^{-i} = \frac{\hfill 1\hfill|}{|\hfill A_1(x)\hfill}
+ \frac{\hfill 1\hfill|}{|\hfill A_2(x)\hfill}
+ \frac{\hfill 1\hfill|}{|\hfill A_3(x)\hfill}
+ \dots
&\mapsto& (A_i(x))_{i=1}^\infty
\end{array}$

We further define $(0^\infty){ \stackrel {G}\longmapsto}
\ 0\ { \stackrel {\cal K}\longmapsto}\ \varepsilon\  \in\  
(\ff_q[x]\backslash \ff_q)^*$,
the empty sequence of (no) polynomials.

The next step gets us back to $\fqo$. We encode the polynomials with positive 
degree by words from $\ff_q^*$, where a polynomial of degree $d\geq 1$ is 
encoded in $2d$ symbols from
$\ff_q$, the first $d$ symbols determining the degree and the second half
determining the coefficients.
Let
$$\Pi_q := \{(a_1,\dots,a_n)\in\ff_q^*\ | \ \exists d\in\nn : n = 2d,
a_1 = \dots = a_{d-1} = 0, a_d \neq 0\}$$
be the set of all allowed encodings of polynomials. Then we define the
encoding function $\pi$ and the subdivision into  degree and coefficient part
as:
$$
\begin{array}{llllll}
\pi\colon \ff_q[x]\backslash\ff_q\ \  &\to&\ \ \Pi_q \subset \ff_q^*,
&\sum_{i=0}^{d}a_ix^i&\mapsto& 0^{d-1}a_d a_{d-1}\dots a_1 a_0\in
\ff_q^{2d}\\
\pi_D \colon \ff_q[x]\backslash\ff_q\ \ &\to&\ \ \ff_q^*,
&\sum_{i=0}^d a_ix^i&\mapsto& 0^{d-1}a_d\in \ff_q^{d}\\
\pi_C \colon \ff_q[x]\backslash\ff_q\ \ &\to&\ \ \ff_q^*,
&\sum_{i=0}^d a_ix^i&\mapsto& a_{d-1}\dots a_1a_0\in \ff_q^{d}
\end{array}$$
($a_d \neq 0$)

Hence $\pi$ induces a function $\pi^\infty$ on the set
$(\ff_q[x]\backslash\ff_q)^* \cup (\ff_q[x]\backslash\ff_q)^\infty$ of 
finite or infinite sequences of polynomials with positive degree as

$\begin{array}{llllll}
\pi^\infty\colon (\ff_q[x]\backslash \ff_q)^*&\to&\ff_q^\infty,
&(A_i)_{i=1}^k&\mapsto& \pi(A_1)|\dots |\pi(A_k)|0^\infty\\
\pi^\infty\colon (\ff_q[x]\backslash \ff_q)^\infty&\to&\ff_q^\infty,
&(A_i)_{i=1}^k&\mapsto& \pi(A_1)|\pi(A_2)|\dots\\
\end{array}$

where $|$ indicates concatenation of elements from $\ff_q^*$.

In a similar way, $\pi^\infty_D$ and $\pi_C^\infty$ are built up from the
$\pi_D$ and $\pi_C$.

The set $\Pi_q' := \Pi_q \cup \{0^\infty\}$ is a complete prefix
code for $\ff_q^\infty$, {\it i.e.}~every sequence $a\in\ff_q^\infty$
can be decomposed in exactly one way into elements from $\Pi_q'$
and hence $\pi^{-\infty}\colon\fqo\to (\ff_q[x]\backslash \ff_q)^* \cup
(\ff_q[x]\backslash \ff_q)^\infty$ is bijective.
\\\\
{\it B.\ The Continued Fraction Operator $\kk$: $\fqo \to \fqo$}

We thus map the continued fraction expansion of a generating function
back into the space $\ff_q^\infty$ and define the {\it Continued Fraction 
Operator} ${\kk}$ on $\ff_q^\infty$ as
$$\kk\colon \fqo\to\fqo,\ \
\kk= \pi^\infty\circ{\cal K}\circ G.$$

The subdivision of the encodings $\pi$ and $\pi^\infty$ into the parts
$\pi_D, \pi_C$ and $\pi^\infty_D, \pi^\infty_C$, resp., define the degree and
the coefficient part of $\kk$ as
$$\kd = \pi^\infty_D \circ {\cal K} \circ G,$$
$$\kc = \pi^\infty_C \circ {\cal K} \circ G.$$

We define ${\kk}$ for finite (prefix) words by
$$\forall a\in
\ff_q^*:\ \kk(a) := \kk(a|0^\infty)_{i=1}^{| a|}$$
(we will see in Theorem 5 that the continuation  $0^\infty$ after $ a$
is irrelevant)
and we obtain the inverse operator as
$${\kk}^{-1} = G^{-1} \circ {\cal K}^{-1} \circ \pi^{-\infty}.$$

($\kd$ and $\kc$ are not injective and thus ${\kd}^{-1}$, ${\kc}^{-1}$ do not make sense)

We give two examples  for ${\kk}, {\kk}^{-1}$:

$(i)$ Let $a=(a_i)_{i=1}^\infty = 1(110)^\infty\in\ff_2^\infty$, then
\begin{eqnarray*}
G(a)(x)&=& x^{-1}+ x^{-2}+ x^{-3}+ x^{-5}+ x^{-6}+ x^{-8}+ x^{-9}+\dots\\
&=&\frac{1}{x} + \frac{x^{-2}+x^{-3}}{1+x^{-3}}
=\frac{x^2+1}{x^3+x^2+x}\cdot \frac{x+1}{x+1} = \frac{1}{x+1+\frac{1}{x^2+1}},
\end{eqnarray*}
from $x^3+x^2+x=(x^2+1)(x+1)+1$.
Thus ${\cal K}(G(a)) = (x+1, x^2+1)\in \ff_2[x]^2$ and
$\kk(a) = \pi^\infty\circ{\cal K}\circ G(a) = 1101010^\infty\in\fqo$,
where $11= \pi(x+1)$, $0101=\pi(x^2+1)$.

$(ii)$ Let  $a = 11011001001011\dots \in \ff_2^\infty$. By repeatedly
taking the integral part and inverting, we obtain
$$G(a)(x) = \frac{\hfill 1 \hfill |}{|x+1}+\frac{\hfill 1 \hfill |}{|x}
+\frac{\hfill 1 \hfill |}{|x^3+x+1}+ \frac{\hfill 1 \hfill |}{|x+1}+\dots$$
The encodings of the partial denominators are $\pi(x+1)=11$, $\pi(x)=10$,
$\pi(x^3+x+1)=001011$, and again $\pi(x+1)=11$, hence
${\kk}(a) = 111000101111\dots$

Let $A_i$  be the partial denominators of $G(a)$. Then we iteratively obtain
{\it convergents} $\frac{P_i}{Q_i}$ to $G(a)$ using
$ P_i := A_i\cdot P_{i-1} + P_{i-2}$ and
$ Q_i := A_i\cdot Q_{i-1} +Q_{i-2}$ with  the initial conditions
$P_{-2} = Q_{-1} = 0, P_{-1} = Q_{-2} = 1$.

The recursion for the $P_i, Q_i$ leads us to 

{\it Theorem $1$} \ \
{\it $P_n\cdot Q_{n-1} - P_{n-1}\cdot Q_n=(-1)^{n-1}$ for $n \geq -1$ }

\bw\
By induction on $n$. See also  \cite[\S 6]{Per}.\hfill $\Box$

The next theorem gives a bound for the precision of the approximation of
$G(a)$ by $\frac{P_k}{Q_k}$.

{\it Theorem $2$} \ \
{\it
Let $\frac{P_k}{Q_k}$ be a convergent to $G\in\ser$ with $G\neq  \frac{P_k}{Q_k}$.
Then

$(i)$ $|{\de}G-\frac{P_k}{Q_k}|{\de} = -|{\de}Q_k|{\de} - |{\de}Q_{k+1}|{\de} < -2\cdot |{\de}Q_k|{\de}$

$(ii)$ For $k\in\nn$ and all $Z,N \in\ff_q[x]$ with
$0 \leq |{\de}N|{\de} < |{\de}Q_{k+1}|{\de}$ we have:
$$\bigg|{\de}G - \frac{Z}{N}\bigg|{\de} \geq \bigg|{\de}G-\frac{P_k}{Q_k}\bigg|{\de} = -|{\de}Q_{k}|{\de}-|{\de}Q_{k+1}|{\de}.$$
}

\bw\ \ \   (compare \cite[p.~69--74]{Mat})\\
$(i)$ Let $\xi_i, A_i$ as before. By induction on $i$ we obtain the equation
$$\forall i\in\nn_0:\ G = \frac{\xi_i\cdot P_{i-1}+P_{i-2}}
{\xi_{i}\cdot Q_{i-1}+Q_{i-2}}$$
and hence
\begin{eqnarray*}
\bigg|{\de}G-\frac{P_i}{Q_i}\bigg|{\de} &=&
\bigg|{\de}\frac{\xi_{i+1}\cdot P_i+P_{i-1}}{\xi_{i+1}\cdot Q_i+Q_{i-1}} -
\frac{P_i}{Q_i}\bigg|{\de}\\
&=&\bigg|{\de}\frac{\xi_{i+1}\cdot (P_iQ_i - P_iQ_i) + (P_{i-1}Q_i - P_iQ_{i-1})}
{\xi_{i+1}\cdot Q_i{}^2 + Q_iQ_{i-1}}\bigg|{\de}\\
&=&|{\de}(-1)^i|{\de}-2|{\de}Q_i|{\de} -|{\de}\xi_{i+1}|{\de}\\
&=& -|{\de}Q_{i+1}|{\de}-|{\de}Q_i|{\de},\\
\end{eqnarray*}
where we have used Theorem 1 and
$|{\de}\xi_{i+1}|{\de} = |{\de}A_{i+1}|{\de} =|{\de}Q_{i+1}|{\de} - |{\de}Q_{i}|{\de}$.

$(ii)$ See \cite[p.221, Th.~B.1]{NrSIAM}.\hfill $\Box$

The approximations $\frac{P_i}{Q_i}$ are called {\it convergents}. Furthermore,
there may occur intermediate results \cite[p.~71]{Mat}, which we shall call
{\it subconvergents}: Let $A_i(x)=\sum_{j=0}^{|{\de}A_i|{\de}} a_j^{(i)}\cdot x^j$ 
be a partial denominator, that is
$P_i = A_i\cdot P_{i-1}+P_{i-2}$ and~$Q_i = A_i\cdot Q_{i-1}+Q_{i-2}$, resp.,
a (main) numerator and denominator, resp. Then for each
$k=|{\de}A_i|{\de},|{\de}A_i|{\de}-1,\dots,2,1$ we can define an auxiliary partial denominator
$A_i^{(k)}(x) := \sum_{j=k}^{|{\de}A_i|{\de}} a_j^{(i)}\cdot x^j$ that in turn defines
a subconvergent
$$\frac{P_i^{(k)}(x)}{Q_i^{(k)}(x)} := \frac{A_i^{(k)}\cdot P_{i-1}+P_{i-2}}
{A_i^{(k)}\cdot Q_{i-1}+Q_{i-2}}.$$
In the preceeding example,  $A_2(x) = x^2+x+1$ and
$\frac{P_2(x)}{Q_2(x)} = \frac{x^2+x+1}{x^4+x^2}$.
From $A_2^{(2)}(x) = x^2$ and $A_2^{(1)}(x) = x^2+x$ we obtain the
subconvergents

\begin{eqnarray*}
\frac{P_2^{(2)}(x)}{Q_2^{(2)}(x)} &=&
\frac{x^2\cdot1+0}{x^2\cdot(x^2+x+1)+1}=
\frac{x^2}{x^4+x^3+x^2+1}\ \ {\rm and}\\
\frac{P_2^{(1)}(x)}{Q_2^{(1)}(x)} &=&
\frac{(x^2+x)\cdot1+0}{(x^2+x)\cdot(x^2+x+1)+1} =
\frac{x^2+x}{x^4+x+1}.\\
\end{eqnarray*}

We will see in the proof of Theorem 5 that the subconvergents 
$\frac{P_i^{(k)}}{Q_i^{(k)}}$ have
a precision between that of the convergents $\frac{P_{i-1}}{Q_{i-1}}$ and
$\frac{P_i}{Q_i}$. All convergents including the subconvergents will be
obtained during the calculation via Euclid's algorithm.

Of course, Theorem 1 is also valid for subconvergents
$$P_i^{(k)}\cdot Q_{i-1} - P_{i-1} \cdot Q_i^{(k)} = (-1)^{i-1},$$
(again by induction, on $k$, for fixed $i$) 
see also {Carter} \cite[Lemma 4.2.1]{Car}.
\\\\
{\it C.\ \  $\kk$ is an Isometry}

In the sequel, we will see that the $n$--th symbol ${\kk}(a)_n$
depends exactly on $a_1,\dots,a_n$, but not on $a_{n+1},\dots$.
In particular, this means that at the end of $\kk(a)$ for
$a\in\ff_q^*$ we might have an incomplete encoding. This then describes
exactly what is known at that position about the partial denominator.

In the next theorem, we show that $\kk^{-1}$ is an isometry on $\fqo$ that is,
two inputs differing for the first time in position $n$ lead to outputs also 
differing here, but not before.

{\it Proposition $3$} {\it ``The Ultrametric Square''}\ \ \ 
{\it
Let $P,Q,R,S\in\fqo$ be four ``points'' with $|P-Q|, |R-S| < |P-R|$.
Then $|Q-S|=|P-R|$.
}

\bw\ \ 
By assumption and the ultrametric inequality 
$|P| \neq |Q|\Rightarrow |P\pm Q| = \max\{|P|,|Q|\}$, we have
$|P-S| = |(P-R)+(R-S)|= \max\{|P-R|,|R-S| \} = |P-R|$. By the same 
reasoning $|Q-S| = |(Q-P)+(P-S))| = |P-R|$.~\hfill~$\Box$

{\it Lemma $4$}\ \ 
{\it
Let $c\in \fqo$ with $\kk^{-1}(c)=a$, $n\in\nn$, and 
$d\in\fqo$ with $d_i=c_i$ for $i\neq n$, $d_n\in\ff_q\backslash\{c_n\}$, and
$b=\kk^{-1}(d)\in\fqo$.
Then $|a-b|=-n$.
}

\bw\ \ 
Let the continued fraction of $G(a)$ be  $G(a) = 
\frac{\hfill 1\hfill|}{|\hfill A_1(x)\hfill}
+ \frac{\hfill 1\hfill|}{|\hfill A_2(x)\hfill}
+ \frac{\hfill 1\hfill|}{|\hfill A_3(x)\hfill} +\dots$. Then  we have 
$c= \pi(A_1)|\pi(A_2)|\pi(A_3)|\dots$
Similarly, let
$G(b) = 
\frac{\hfill 1\hfill|}{|\hfill B_1(x)\hfill}
+ \frac{\hfill 1\hfill|}{|\hfill B_2(x)\hfill}
+ \frac{\hfill 1\hfill|}{|\hfill B_3(x)\hfill} +\dots$ and thus 
$d= \pi(B_1)|\pi(B_2)|\pi(B_3)|\dots$. 
Now let $k\in\nn$ be such that $A_i=B_i$ for $i<k$ and $A_k\neq B_k$.
This implies $|Q_{k-1}|+|Q_{k}|\leq n \leq 2|Q_k|$ ($c_n$ is $lc(A_k)$ or 
part of $\pi_C(A_k)$).
W.l.o.g.~let $|A_k|\leq |B_k|$. We have
$\big|G(a)-\frac{P_k}{Q_k}\big| =-|Q_k|-|Q_{k+1}|< -2|Q_k|\leq -n$ by 
Theorem 2(i). We 
consider the ultrametric square made up of $G(a), G(b)$ and their convergents 
$\frac{P_k}{Q_k}, \frac{\tilde P_k}{\tilde Q_k}$.
First we treat the case  $|A_k|=|B_k|$.
Then $B_k = A_k +(d_n-c_n)x^d$ with $g=2|Q_k|-n$ ($g$ more symbols after $c_n$
until the end of $\pi(A_k)$).

Then
$\Big|G(b)-\frac{\tilde P_k}{\tilde Q_k}\Big|
=\Big|G(b)-\frac{P_k+x^g P_{k-1}}{Q_k+x^g Q_{k-1}}\Big| 
= -|\tilde Q_k|-|\tilde Q_{k+1}| < -2|\tilde Q_k| \leq -n$ and
$\Big|\frac{P_k+x^g P_{k-1}}{Q_k+x^g Q_{k-1}}-\frac{P_k}{Q_k}\Big| 
=\Big|\frac{P_kQ_k+x^gP_{k-1}Q_k-P_kQ_k-x^gP_kQ_{k-1}}
{Q_k(Q_k+x^gQ_{k-1})}\Big|
=\Big|\frac{x^g(-1)^{k}}{Q_k(Q_k+x^gQ_{k-1})}\Big|
=g-2|Q_k|=-n$.

Altogether we have an ultrametric square where
$\big|G(a)-\frac{P_k}{Q_k}\big| <-n$,
$\big|G(b)-\frac{\tilde P_k}{\tilde Q_k}\big| <-n$,
$\big|\frac{P_k}{Q_k}-\frac{\tilde P_k}{\tilde Q_k}\big| =-n$,
and with Proposition~3 now follows
$$|G(a)-G(b)| = |a-b| = -n = |c-d| = |\kk(a)-\kk(b)|.$$

Let now  $|A_k|<|B_k|$. In this case $c_n = lc(A_k)$, $d_n=0$ and 
$n = |Q_{k-1}|+|Q_k|$.
Then $\Big|G(a)-\frac{P_{k-1}}{Q_{k-1}}| = - |Q_{k-1}| - |Q_{k}|=-n$
and $\Big|G(b)-\frac{P_{k-1}}{Q_{k-1}}| = - |\tilde Q_{k-1}| - |\tilde Q_{k}|
<-n$ and thus as before $G(a)-G(b)|=-n$.\hfill $\Box$

In the sequel we shall say that we are in case A, B, C, resp., if $c_n$ is\\
A) a zero of some $\pi_D$ (in part $(ii)$ of the above example for $n= 5,6$),\\
B) the last coefficient (for $n=2, 4, 10, 12$),\\
C) the leading or some other coefficient except the last 
(for $n=1,3,7\dots 9,11$).

{\it Theorem $5$}\ \ 
{\it
Let $n\in\nn$ and $c,d\in \fqo$ with $d_i=c_i$ for $i< n$, $c_n\neq d_n$.

Let $a=\kk^{-1}(c)\in\fqo$ and $b=\kk^{-1}(d)\in\fqo$. Then
$$|a-b|=|\kk(a)-\kk(b)|=-n.$$
}

\bw\ \ 
We shall change from $(c_i)$ to $(d_i)$ one symbol at a time, using Lemma 4.
Let $d_i^{(k)}
= \Bigg\{
\ba{ll}
d_i,&i \leq k\\
c_i,&i   > k
\ea
$
Then $d^{(0)} = c$ and 
$\lim_{k\to\infty}d^{(k)} = d$.
Let $b^{(k)} = \kk^{-1}(d^{(k)})$. Then $|a-b^{(0)}| = -\infty$,
since $a= b^{(0)}$.
Furthermore, 
$|b^{(i-1)}-b^{(i)}|
= \Bigg\{
\ba{ll}
-\infty,&c_i = d_i\\
-i,&c_i\neq d_i
\ea
$
from $d^{(i-1)} = d^{(i)}$ and by Lemma 4, respectively.

Hence $|a-b^{(n)}|= -n$ (first difference between $c$ and $d$),
and $|b^{(n+k)}-b^{(n+k+1)}| \leq -n-k-1< -n$ for $k\in\nn_0$.

By the ultrametric equality, we thus have 
$|a-b^{(n+k)}| = |(a-b^{(n)}) + (b^{(n)}-b^{(n+1)})+\dots 
+(b^{(n+k-1)}-b^{(n+k)}| = -n$ for all $k\in \nn$, 
that is after the first impact of changing $c_n$ into $d_n$, 
the further changes are ``absorbed'' by the ultrametric. 
In the limit $k\to\infty$, we obtain
$|a-b| = -n = |\kk(a) - \kk(b)|$. \hfill $\Box$
\\\\
In summary, the first $n$ symbols of $a$ determine the first $n$ symbols of
$\kk(a)$ and vice versa.
At each step, the resulting symbols from $\ff_q$ encode {\it all} information
available at that point about the continued fraction expansion:
In case B, we just have the complete last determined partial denominator, 
in case A, the next partial denominator $A_k$ will have degree at least 
$n-2|Q_{k-1}|$ and that is all we can say at the moment, 
and in case C, the missing coefficients of the current partial denominator 
are as yet unknown and {\it all} 
possible values will be assumed for a suitable continuation of $a$.

We have thus obtained the fundamental result of the paper:

{\it Theorem $6$}\ \

{\it
The function $\kk\colon \fqo\to \fqo$ is an isometry on $\fqo$, for every 
finite field~$\ff_q$.
}

{\it Remark}\ \
We call the
domain of $\kk$ (range of ${\kk}^{-1}$) {\it coefficient space } and the
range of $\kk$ (domain of ${\kk}^{-1}$)  {\it discrepancy space}
(as in the {Berlekamp--Massey}--Algorithm).
\\\\\\\\
\centerline{\sc III. Linear  Complexity,\ \ Euclid's Algorithm}
\\\\
We recall the notion of linear complexity $L$, give the connection between
$\kk$ and $L$, and adapt the Berlekamp--Massey algorithm to produce precisely 
$b=\kk(a)$ as discrepancy sequence on input $a$.
Furthermore, we describe an alternative method, due to Niederreiter and the 
author, to calculate $\kk(a)$ by means of the shift commutator 
$[\kk^{-1},\sigma]$.
\\\\
{\it A.\ Linear Complexity}

We define the {\it linear complexity profile} of a sequence $a\in\ff_q^\infty$
as $L_a\colon\nn_0\to \nn_0$ with $L_a(0)=0$ and for $n\geq 1$ let
$L_a(n) = \sum_{i=1}^k |{\de}A_i|{\de}$, where $A_k$ is the last partial denominator
whose leading coefficient is encoded in $\kk(a)_{i,i=1\dots n}$.

The sequence $(L_a(n))_{n\geq 0}$ is called the {\it linear complexity
profile} of $a$. $(L_a(n))$ is monotonously increasing and jumps,
$L_a(i) > L_a(i-1)$, where
$\kk(a)_i$ encodes a leading coefficient. 
We next show that the usual definition in linear feedback shift
register (LFSR) theory is equivalent:

{\it Theorem $7$}\ \
{\it $L_a(n)$ denotes the length of a shortest LFSR, that produces
$a_1\dots a_n$.}

\bw\
Given the formal power series $G(a)$ with the convergent numerators and
denominators $(P_k)$ and $(Q_k)$, resp., we consider the LFSR with normalized
feedback polynomial $(lc(Q_k))^{-1}\cdot Q_k$ of length
$|{\de}Q_k|{\de}=\sum_{i=1}^k|{\de}A_i|{\de} = L_a(n)$.
This LFSR produces, for a suitable initial content, a sequence $b$ with
$G(b)=\frac{P_k}{Q_k}$.
From $|{\de}G(a)-\frac{P_k}{Q_k}|{\de} = -|{\de}Q_{k+1}|{\de} -|{\de}Q_{k}|{\de}<-n$
now follows that this LFSR will produce $a$ (at least) up to $a_n$.
On the other hand,
$$\Big|{\de}G(a)-\frac{P_{k-1}}{Q_{k-1}}\Big|{\de}
= -|{\de}Q_{k}|{\de} -|{\de}Q_{k-1}|{\de} = -2|{\de}Q_{k-1}|{\de}-|{\de}A_k|{\de} \geq -n,$$
since the leading coefficient of $A_k$ lies in $\kk(a)_{i,i=1\dots n}$.
Hence the previous convergent $\frac{P_{k-1}}{Q_{k-1}}$ is {\it not}
sufficiently precise and by Theorem 2$(ii)$ also no intermediate LFSR length 
will do.~\hfill~$\Box$
\\\\
We see that typically $2\cdot L_a(n) \approx n$, since with
$L_a(n) =  \sum_{i=1}^k |{\de}A_i|{\de}$ we need just $2\cdot L_a(n)$ symbols to
encode $A_1,\dots,A_k$, hence the deviation from the sequence length $n$
consists only in the length of the last incomplete encoding of a partial
denominator (missing coefficients in case C of Lemma 4,
exceeding zeroes in case A). Hence we define:

Given a sequence $a$ with linear complexity profile $(L_a(n))$ we
define the {\it linear complexity deviation} of $a$ at $n$ as
$$m_{a}(n) := 2\cdot L_a(n) - n\in\zz,\ n\geq 0.$$

{\it Remark} \
Comparing with the three cases of Lemma 4, we have:\\
$m_a(n) < 0$ in case A, $m_a(n) = 0$ in case B, $m_a(n) > 0$ in case C.
\\\\
{\it B.\ Euclid's Algorithm}

{\it History}\ \ Euclid (``Elements'', 300 BC) invents his algorithm to 
calculate the $gcd$ of two natural numbers. Lagrange (1770 AD) has our 
Theorems 1 and 2, all this over the reals. 

{Berlekamp} \cite{Ber} describes in 1967 a decoding method for BCH--codes
over arbitrary finite fields. In 1969, {Massey} \cite{Mas} uses this 
method to obtain a shortest LFSR producing a given sequence.
In 1979 {Welch} and {Scholtz}  \cite{WS} make the connection with
continued fractions and they show that the {\it main} convergents appear
as polynomials in the BMA.

In 1987 Dornstetter \cite{Dorn} gives the precise equivalence between the
algorithms of Berlekamp and Euclid, already mentioning the subconvergents.

During 1988 -- 1991 {Niederreiter} 
\cite{Nr87}\cite{Nr88}\cite{Nr89}\cite{Nr90}
as well as {Dai} and Zeng  \cite{Dai} then show the detailed connection 
between linear
complexity $(L_a)$ and the continued fraction expansion of $G(a)$:
A jump by $k$ in the profile corresponds to a partial
denominator of degree $k$.
Furthermore, in \cite{Dai} a connection between subconvergents and 
discrepancies is given for the case $\ff_2$.
\\\\
Before giving the implementation in a form that exactly delivers $b=\kk(a)$ 
as discrepancy sequence, we shall recompile the changes necessary in comparison
with \cite{Mas}:\\
$(i)$ The main convergents have to start with
$P_{-2}=0,$ $ P_{-1} = 1,$ $ Q_{-2}=1,$ $Q_{-1} = 0$ as is the case over $\rr$ 
\cite{Per}. From $A_0=0$ we also have  $P_0=0,$ $ Q_{0}= 1$,\\
$(ii)$ we use the feedback polynomial, {\it not} its reciprocal, the 
``connection polynomial'', and \\
$(iii)$ the feedback polynomial will {\it not} be normalized.

In this way we will obtain all partial denominators and all convergents
and subconvergents and the discrepancy sequence {\it is} the encoding 
$\kk(a)$ of the partial denominators.
\\\\
{\it Euclid--Lagrange--Berlekamp--Massey--Dornstetter--Algorithm}
\begin{tabbing}
DO D\=DO D\=DO D\=DO D\=DO D\=DO D\=DO D\=DO D\=\kill
START\\
Input a(1,...) // elements of $\ff_q$\\
$P = 0,\ AP = 1$\ // numerator\\
$Q=1,$ $AQ=0,$\ // here Q=$Q_0$, AQ = $Q_{-1}$ as initial values\\
$d=0,$ $m=0,$ $j=0,$ $r=-1$\\
$// loop\ invariant:  m=2\cdot d - j,$ $ d = |{\de}Q|{\de}$\\
DO $j=1,\dots$\\
\>$b(j) = \sum_{i=0}^d Q(i)\cdot a(j+i-d)$\ \
$// {\rm equals}\ \sum Q(d-i)\cdot a(j-i)$\\
\>CASE $(b(j),m)$ IS\\
\>\>$(0,\ \cdot\ ):$\\
\>\>\>$m = m-1$\\
\>\>$(\neq 0,>0):$\\
\>\>\>$m = m-1$\\
\>\>\>$\tilde b = \overline r\cdot b(j)$\\
\>\>\>$P = P+\tilde b \cdot x^m\cdot AP$\ // numerator\\
\>\>\>$Q = Q+\tilde b \cdot x^m\cdot AQ$\\
\>\>$(\neq 0,\leq 0):$\\
\>\>\>$m = -(m-1)$\\
\>\>\>$\tilde b = r$\\
\>\>\>$r=b(j)$\\
\>\>\>$\overline r = -r^{-1}$\\
\>\>\>$\tilde b = \tilde b \cdot \overline r$\\
\>\>\>$P_{tmp} = AP$, \ $AP = P$, \ $P = P_{tmp}$\ // numerator\\
\>\>\>$P = P+\tilde b \cdot x^m\cdot AP$\ // numerator\\
\>\>\>$Q_{tmp} = AQ$, \ $AQ = Q$, \ $Q = Q_{tmp}$\\
\>\>\>$Q = Q+\tilde b \cdot x^m\cdot AQ$\\
\>\>\>$d = d+m$\\
\>END CASE\\
\>$// loop\ invariant:  m=2\cdot d-j,$ $d = |{\de}Q|{\de}$\\
\>Output b(j), P, Q\\
END DO
\end{tabbing}
$P/Q$ is the actual (sub)convergent, $AP/AQ$ the previous convergent,
 $(b(j)) = \kk (a(j))$. The lines with
comment  ``//numerator'' are not necessary to obtain $(b_j)$.
For $\ff_2$ we have $\tilde b = b(j)$, $b(j)\neq 0 \Rightarrow b(j)=1$, and
$r=\overline r = 1$, which simplifies the program.
\\\\
{\it C.\ The Shift Commutator of $\kk$}

Another method to  actually compute $\kk$ uses the shift commutator
$$[\kk^{-1},\sigma] = \kk\circ \sigma^{-1}\circ \kk^{-1}\circ\sigma.$$
Let some $w = \kk(v) $ be given, {\it e.g.}~$0^\infty = \kk(0^\infty)$,
then $\kk(av) = [\kk^{-1},\sigma](aw)$  for $a \in\ff_q$, since
$$aw
\ {\stackrel \sigma \longrightarrow }\ w
\ {\stackrel {\kk^{-1}} \longrightarrow }\ v
\ {\stackrel {\sigma_a^{-1}} \longrightarrow }\ av
\ {\stackrel \kk \longrightarrow }\ \kk(av)
= [\kk^{-1},\sigma](aw)$$

$[\kk^{-1},\sigma]$ can be computed by a transducer with finite state space
and an up-down- counter in amortized linear time (for a detailed description
see \cite{NV3} for the case $\ff_2$ and \cite{NV4} for general finite fields).
Hence we can compute $\kk$ in amortized quadratic time by repeated
application of $[\kk^{-1},\sigma]$ via the above formula. We thereby obtain
with no additional cost {\it all} continued fractions of {\it all} shifted
sequences $(a_1,\dots,a_n),$  $(a_2,\dots,a_n)$ $\dots$  $(a_{n-1},a_n)$
as well.
\\\\\\
\centerline{\sc IV. Consequences for Linear and Jump Complexity  
of $\kk$ being Isometry}
\\\\
We derive the partition of $\kk$ into $\kd$ and $\kc$ from the behaviour of 
the algorithm in III.B. Furthermore, we derive from $\kd$ the linear and 
jump complexity profiles and some combinatorial results.
\\\\
{\it A. \ Translation Theorem}

{\it Theorem $8$}\ \
{\it
For every length  $n\in\nn$ and every sequence prefix $a \in \ff_q^n$
we have:

\begin{tabular}{rlll}
$(i)$&$m_{a}(n) > 0:$ &
$m_{a|a_{n+1}}(n+1) = m_{a}(n) - 1$.\\
$(ii)$&$m_{a}(n)\leq 0:$\\
&$\exists_1 \alpha \in \ff_q :
$&$m_{a | \alpha}(n+1) = m_{a}(n)-1$,\\
&$\forall\ \alpha'\neq \alpha :$&$m_{a | \alpha'}(n+1) =
1 - m_{a}(n)$.\\
$(iii)$&$L_a(n) > n/2:$ &$L_{a|a_{n+1}}(n+1) = L_a(n)$.\\
$(iv)$&$L_a(n) \leq n/2: $&\\
&$\exists_1 a \in \ff_q : $&$L_{a | \alpha}(n+1) = L_a(n)$,\\
&$\forall\ \alpha'\neq \alpha : $&$L_{a | \alpha'}(n+1) = n+1 - L_a(n)$.\\
\end{tabular}
}

(In $\ff_2$ obviously 
$\forall \alpha'\neq \alpha \Leftrightarrow \alpha' = \alpha+1$)

\bw\
$(i)$ Since $m_a(n)>0$, when running the algorithm from III.B, only the cases
$(b,m)=(0,\cdot)$ or $(b,m) = (\neq 0,>0)$ may appear. Hence $m(n+1) = m(n)-1$.

$(ii)$ For exactly one choice $\alpha\in\ff_q$ as $a_{n+1}$ we have $b_{n+1}=0$,
hence the case $(b,m) = (0,\cdot)$ and $m(n+1) = m(n) - 1$.
All other  $\alpha'\neq \alpha$ lead to $b_{n+1}\neq 0$ and 
$(b,m) = (\neq 0,\leq 0)$, thus $m(n+1) = 1 -m(n)$.

$(iii), (iv)$ are  equivalent to $(i)$, $(ii)$ by the definition of $m_a$,
see also \cite[p.~34]{Rup1}.~\hfill~$\Box$

For the following theorem and again in Section V, we need a measure on
$\fqo$. We define the measure $\mu$ on  $\ff_q$ as equidistribution,
$\mu(a) = \frac{1}{q},\ \forall \ a\in\ff_q$. Taking the infinite 
product Haar measure of $\mu$, we obtain the measure  
$\mu^\infty$ on $\ff_q^\infty$. 
A set $A=\{a\in\ff_q^\infty\ |\ a_i=b_i, 1\leq i\leq k\}$ for fixed 
$b_i\in\ff_q$, is called a  {\it cylinder set} and it has measure 
$\mu^\infty(A)=q^{-k}$.
The set of ultimately periodic (rational) sequences in $\ff_q^\infty$ is
countable and thus has measure zero.

{\it  Theorem $9$}\ \  {\it Translation Theorem}\ \\ 
{\it
Let ${\alpha} = (\alpha_1,\dots,\alpha_k)$ and
${\beta} = (\beta_1,\dots,\beta_l)$ be two sequences from
$\ff_q^*$ with
$m_{\alpha}(k) = m_{\beta}(l).$
Furthermore, let $A=\alpha|\fqo$ and $B=\beta|\fqo$ be the cylinder sets of 
the sequences in $\ff_q^\infty$ starting with $\alpha$ or~$ \beta$, resp.
Then we have for all $t \in \nn_0$ and for all $d \in\zz$:

$\begin{array}{lllllllllll}
(i)\ \ &|\ \{ a&\in \ff_q^{k+t}&|\ a_i &= \alpha_i,
&i \leq k,&m_a(k+t)&=d \}\ |\vspace{2 mm}\\
\ \ =&|\ \{ b&\in \ff_q^{l+t}&|\ b_i &= \beta_i,
&i \leq l,&m_{b}(l+t)&=d \}\ |\\
\end{array}$

$$(ii)\ \   \frac{\mu^\infty(\{a \in A\y|\y m_a(k+t) = d\})}{\mu^\infty(A)}
\ =\  \frac{\mu^\infty(\{b \in B\y|\y m_b(l+t) = d\})}{\mu^\infty(B)}.$$

In other words:\
The distribution of deviations  $m$ after a given prefix depends only on the
$m$ at the end of that prefix; it does not depend on the length or the
particular symbols of this prefix.
}

\bw

$(i)$ By induction on $t$: For $t=0$ both sets contain exactly one element for
$d=m_{\alpha}(k)$  (which is $\alpha$
resp.~$\beta$ itself) and for $d\neq m_{\alpha}(k)$ both sets
are empty. The step from $t$ to $t+1$ follows from Theorem 8$(iii,iv)$.

$(ii)$ This is just a measure theoretic reformulation of the result in $(i)$.
\hfill$\Box$

{\it Corollary $10$}\ \
{\it
Let  $m_s(2k) = 0$ for a se\-quence $s$ of length $2k$  $($thus
$L_s(2k) = k)$.
Then the distribution of linear complexity de\-viations
on the cylin\-der set  $s|\ff_q^\infty$ equals that on $\ff_q^\infty$.
}

{\it Example }\
Extending $s=\varepsilon$ (empty sequence) and $s=10$, resp.,
by 3 bits:

{\small{
\begin{tabular}{ccc|r|ccccc|r}
$s_1$&$s_2$&$s_3$&$m(3)$&$s_1$&$s_2$&$s_3$&$s_4$&$s_5$&$m(5)$\\
0&0&0&$-$3\y\y & 1&0& 0&0&0&$-$3\y\y\\
0&0&1& 3\y\y& 1&0& 0&0&1   & 3\y\y\\
0&1&0& 1\y\y & 1&0& 0&1&0  & 1\y\y\\
0&1&1& 1\y\y & 1&0& 0&1&1  & 1\y\y\\
1&0&0& $-$1\y\y & 1&0&1&0&0& 1\y\y\\
1&0&1& 1\y\y & 1&0& 1&0&1  & $-$1\y\y\\
1&1&0& 1\y\y & 1&0& 1&1&0  & $-$1\y\y\\
1&1&1& $-$1\y\y &1&0& 1&1&1& 1\y\y\\
\end{tabular}
}}
\\\\
{\it B.\ Partition of $\kk$ into $\kd$ and $\kc$}

We need a  slight modification of $m$ to be able to relate the positions
in $\kk$ to those in $\kd, \kc$. Let 
$$m'(n) = m(n-1)-1, n>0\mbox{\rm\  and\ } m'(0) = 0$$ 
($m'$ is equal to $m$, except for
the positions of the leading coefficients in $\kk$, since here $m(n) = d$,
but $m'(n) = -d$, with  $d$ the degree of the current partial denominator).

{\it Theorem $11$}\ \
{\it
Using $m'$ we have the following connection between
the linear complexity deviation and the distribution of $\kk$ onto $\kd$
and $\kc$:\\
$m'(n) < 0:    \kk(a)_n = \kd(a)_{(n - m'(n))/2}$\\
$m'(n) \geq 0: \kk(a)_n = \kc(a)_{(n - m'(n))/2}$
}

\bw\
By induction on the codings $\pi$: Let some $\pi$ end at the (even) position
$2k$. Then $\kd$ and $\kc$ both contain $k$ of the $2k$ symbols
$\kk(a)_{i,i=1\dots 2k}$ up to that position and $\kk(a)_{2k}=\kc(a)_k$.
Also, before the {\it first} coding we have $k=0$ and
$\kd=\kc=\varepsilon$ empty up to now.

The next $d$ symbols ($d$ the degree of the next possibly incomplete next
partial denominator or the length  of the terminating  zero run) lie in $\pi_D$
and yield $m'(n) = m'(2k+i) = -i < 0,\ 1\leq i\leq d$, hence
$\kk(a)_{2k+i}$ $= \kd(a)_{k+i} $ $= \kd(a)_{(2k+i-(-i))/2} $
$= \kd(a)_{(n - m'(n))/2}$  for $1\leq i \leq d$.

For $2k+d<n\leq 2k+2d$, now follows the $\kc$ part ($\kk(a)_{2k+d}$ was a 
leading coefficient), and
the positions $n=2k+d+i, 1\leq i\leq d$ (as far as present) have
$m'(n) = m'(2k+d+i) = d-i\geq 0$ and
$\kk(a)_{2k+d+i}= \kc(a)_{k+i} = \kc(a)_{(2k+d+i-(d-i))/2} =
\kc(a)_{(n - m'(n))/2}$  for $1\leq i \leq d$.~\hfill~$\Box$
\\\\
{\it C.\ $\kd$ and the Linear Complexity Profile}

{\it Theorem $12$}\ \
{\it
The distribution of $m_a(t)$ on the set $\ff_q^t$ of all prefixes of length
$t>0$ can be calculated via $\kk$ and~$\kd$, by regarding
$\ff_q^t$ as discrepancy space.

$(i)$ For even $t>0$ there are $(q-1)\cdot q^{t-1}$ sequences with $m(t)=0$.

$(ii)$ For $0 > m(t)\equiv t\ {\rm mod}\ 2$ with $|m(t)| \leq t$ there are
$(q-1)\cdot q^{t-1+m(t)}$  sequences with $m(t) > -t$ and one sequence
with $m(t) = -t$.

$(iii)$ For $0 < m(t)\equiv t\ {\rm mod}\ 2$ with $|m(t)| \leq t$ there are
$(q-1)\cdot q^{t-m(t)}$  sequences with $m(t) = t$.
}

\bw\
The set  $\ff_q^t$ of all sequence prefixes of length $t$ yields also
$\ff_q^t$ as set of all discrepancy prefixes, since $\kk$
is an isometry.

$(i)$ Since $m(t)=0$, both $\kd$ and $\kc$ have $\frac{t}{2}$ symbols
up to now, which (only) requires $\kd(a)_{t/2}\neq 0$ to be a 
leading coefficient. The other $t-1$ symbols are arbitrary, hence we get
$(q-1)\cdot q^{t-1}$ possible sequences.

$(ii)$ With $m(t)<0$, $m(t)=m'(t)$.
For $m(t)\neq -t,$ $\kk(a)_{i,i=t-m(t)-1\dots t} =$
$\kd(a)_{i,i=(t+m(t))/2\dots (t-m(t))/2}$ $=lc 0^{|m(t)|}$, for some $lc\neq 0$,
thus $(q-1)\cdot 1^{|m(t)|}\cdot q^{t-|m(t)|-1}$ cases,
and
for $m(t)=-t,$ $\kk(a)_{i,i=1\dots t} = \kd(a)_{i,i=1\dots t} =0^t$ (1 case).

$(iii)$ Since $m(t)>0$, the last $\kd$ has encoded a polynomial of degree 
$m(t)$ or higher and thus ended in $0^{m(t)-1}\alpha, \alpha\neq 0$. 
The initial part of $\kd$ and all of $\kc$ are irrelevant for this $m(t)$,
and we have $1^{m(t)-1}\cdot (q-1)\cdot q^{t-m(t)}$ possible 
sequences.~\hfill~$\Box$

{\it Corollary $13$}\ \    ({\it see also Gustavson {\rm \cite{Gus})}}\\
{\it
Let $N(t,m(t))$ be the number of sequences with a given length $t$ and
value $m(t)$. Then for all $t\in \nn_0$ and $m(t)$ with $-t\leq m(t)\leq t$
and $m(t)\equiv t  \mod 2$
$$N(t,m(t)) =\Bigg\{
\ba{ll}
1,&m(t) = -t\\
(q-1)\cdot q^{t-|m(t)-\frac{1}{2}| - \frac{1}{2}},&{\it otherwise}\hspace*{3 cm}\Box\\
\ea$$
}
\\
The following theorem  will finally be the foundation for Theorem~16 
and the following section. We shall see that global statements about the
behaviour of  $m, L $ and $J$  are considered  advantageously in
the discrepancy space. For equidistributed $a$ the resulting  $\kd(a)$
in the discrepancy space are also equidistributed by the following theorem
(with respect to $\kk(a)$ we have equidistribution anyway,
since $\kk$ is isometry).

{\it Theorem $14$}
{\it
$$\forall n \in \nn_0, \forall b \in \ff_q^n:
\ |\ \{a\in\ff_q^{2n}\ |\ \kd(a)_{i,i=1\dots n} = b\}\ | = q^n$$

The  $q^{2n}$ sequences $a$ are thus equidistributed concerning the
$\kd$ part of their partial denominators.
}

\bw\
Instead of $a \in\ff_q^{2n}$ we consider the sequence
$a' = \kk(a) \in \ff_q^{2n}$ in the discrepancy space. Since
$\kk$ is an isometry, every $a'\in\ff_q^{2n}$ occurs exactly once for an
$a$ and the theorem is equivalent to
$$\forall n\in\nn_0,\forall b \in\ff_q^n: \ |\ \{a' =
\kk(a)_{i,i=1\dots 2n}\in\ff_q^{2n} \ |\ \kd(a)_{i,i=1\dots n}=b\}\ | = q^n.$$

We thus have to consider only the partition of $\kk(a)$ into the parts
$\kd$ and $\kc$. First of all we have $|\kd(a)|\geq |\kc(a)|$,
such that  $2n$ symbols of $\kk(a)$ include at least  $n$ symbols of
$\kd(a)$.

Now let  $b=\kd(a)$ be given. Then, by Theorem~11 the indices $t_1,\dots,t_n$  
with $\kk(a)_{t_i}= \kd(a)_i=b_i,1 \leq i \leq n$
are fixed, and the other symbols $\kk(a)_s, s \neq t_i,\forall i$,
can be chosen arbitrarily, that is in $q^{n}$ ways, without affecting
$\kd(a)_{i,i=1\dots n}$. This proves the theorem.~\hfill~$\Box$

{\it Theorem $15$}

\nopagebreak

{\it
$(i)$ $\kd(a)_n \neq 0$  if and only if the linear complexity profile
assumes the value~$n$.

$(ii)$ We can infer the whole linear complexity profile already from
$\kd$ alone $($compare~{Wang} 
}
\cite[{\it Th.} $2.4$]{Wan2}).

\bw

$(i)$ The linear complexity profile assumes the value $n$, if one of the sums
$l_k :=\sum_{i=1}^k |{\de}A_i|{\de}$ is equal to $n$. In this case
$\kd$ starts with 
$$0^{|{\de}A_1|{\de}-1}lc_10^{|{\de}A_2|{\de}-1}lc_2\dots 0^{|{\de}A_k|{\de}-1}lc_k,$$
where $lc_i = lc(A_i)\in\ff_q\backslash\{0\}$ are the leading coefficients,
and thus $\kd(a)_n = lc_k\neq 0$.

On the other hand, all indices  $n$ with
$l_k  < n < l_{k+1}$ for some $k$ lead to
$\kd(a)_n = 0$, since this element is in the
$0^{|{\de}A_{k+1}|{\de}-1}$ part.

$(ii)$ By $(i)$ we know from $\kd$, which linear complexities 
 $l_1<l_2<\dots \in\nn$  occur at all, and we put  $l_0=0$.

At the end of each  $\pi$,  $m$ is zero, hence
$\forall i\in\nn_0: m_a(2\cdot l_i)=0$ and $L_a(2\cdot l_i)=l_i$.
For those $n\in\nn$ with $n\neq l_i$ for all $i\in\nn_o$ there is a
$k$ with $2\cdot l_k<n<2\cdot l_{k+1}$ and then
$L_a(n)=\Big\{$
\bt{ll}
$l_k,$&$k < l_k+l_{k+1}$\\
$l_{k+1},$&$k \geq l_k+l_{k+1}$\\
\et
(where $\kk(a)_{l_k+l_{k+1}} = lc(A_{k+1})).$\hfill$\Box$

In the next theorem we shall obtain the number of occurences of values $m(t)$
by counting  strings in $\kd$.

{\it Theorem $16$}

{\it
$(i)$ A linear complexity deviation  $m_a = k \leq 0$ occurs wherever $\kd$
contains a string $\alpha0^{|k|}, \alpha \neq 0$,
or if $a$ begins with $0^{|k|}$.

$(ii)$  A linear complexity deviation  $m_a = k > 0 $ occurs  wherever $\kd$
contains a  string $0^{k-1}\alpha, \alpha \neq 0$.
}

\bw

$(i)$ The case $a=0^{|k|}\dots$ is obvious. When $\kd(a)$ contains an
$\alpha 0^{|k|}$, then  $\alpha$ terminates a $\pi_D$ and $0^{|k|}$ leads to a
complexity deviation $m=k\leq 0$.

On the other hand, by Theorem 9 this $m_a$ must result from a $\pi_D$, 
which begins  with $0^{|k|}$, thus either $\kd=0^{|k|}$ (at the beginning 
of $a$) or $\kd$ contains the string $\alpha 0^{|k|}$.

$(ii)$ A complexity deviation $m_a=k>0$ occurs exactly after a jump
to $k$ or higher, hence in a coding $\pi(A_i) = 0^{d-1}c_dc_{d-1}\dots c_0$ 
with $d \geq k$, at $c_d$.~\hfill~$\Box$
\\\\
{\it D.\ Iterated Application of \kk}

Pseudorandom sequences should avoid easily guessable 
patterns, for instance neither $a$, nor $\kk(a)$ should terminate in 
$0^\infty$.
Since $\kk$ is an isometry, we can apply $\kk$ again on $\kk(a)$ to obtain
$\kk^2(a)$ which ends in $0^\infty$ exactly for quadratic--algebraic $a$
(with rational $\kk(a)$) and also is far from random.
In fact, for {\it every} exponent $k$, the sequence $\kk^k(a)$ should be
well--behaved, that is look random.

Hence, defining $\kk^\infty$ as $(\kk^\infty)(a)_k := (\kk^k)(a)_k$, 
also $\kk^\infty$ should behave well.

{\it Conjecture}\ \
{\it 
Whenever $\kk^i(a)$ is rational for some $i$, $\kk^\infty(a)$ is 
algebraic.
}

Then by a result of Christol {\it et al.} {\rm \cite{CKMR}},  $\kk^\infty(a)$ 
has  finite tree complexity {\rm \cite{NVTree}},
which can serve as a means to determine this $a$ as nonrandom.
\\\\
{\it E.\ Jump Complexity}

We represent the jump complexity $J_a(t)$, as introduced by  {Carter}
\cite{Car} and {Wang} \cite{Wan1} via $\kk$ by the number of nonzero
symbols in $\kd(a)$. 
The jump heights then will be lengths of zero runs
in $\kd(a)$. This prepares the global results on  $J$,
$m$ and $L$ in the next section.
The {\it Jump Complexity}
$$J_a(n) := |\ \{k\ | \ 1\leq k\leq n \ \land\ L_a(k-1)<L_a(k)\}\ |,
\ n\in\nn_0$$
counts the number of jumps in the linear complexity profile of the sequence
$(a_1,\dots,a_n)$  (see {Carter} \cite{Car}).

{\it Theorem $17$}  { Jump complexity and the discrepancy space}

{\it
$(i)$ The distribution of jump heights corresponds to the distribution of
zero runs in $\kd:$ every zero run of length $k-1, k \geq 1$
corresponds to a jump by~$k$.

$(ii)$ The average distribution of jump heights on  $\ff_q^\infty$
can be modelled by the distribution of zero runs in $\kd$.
Hence a jump by $k$ occurs with probability $p(k) = (q-1)/ q^{k}$.

$(iii)$ $J_a(2\cdot n) =
|\{i\ |\ \kd(a)_i\neq 0, 1 \leq i\leq n\}|+ \delta$,
with $\delta = 1$ for  $m_a(2n) > 0$ and
$\delta = 0$ otherwise.

$(iv)$ We set
\[J'_a(2n) =\Bigg\{
\ba{ll}
J_a(2n),  &{\rm if}\ m_a(2n)\leq 0\\
J_a(2n)-1,&{\rm if}\ m_a(2n)> 0\\
\ea\]

Then
$|\{a\in\ff_q^{2n}\ |\ J'_a(2n) = j\}|$
$= |\{b\in\ff_q^{n}\ |\ b\ $
$ \mbox{\it contains\ exactly\  j\  nonzeroes}\}|$
$={n \choose j} \cdot(q-1)^j.$
}

\bw

$(i)$ Every jump by  $d$ corresponds to a  $\pi_D$ of 
$0^{d-1}\alpha, \alpha \in \ff_q\backslash\{0\}$
that is a sequence $\beta 0^{d-1}\alpha,\ \alpha,\beta\in\ff_q\backslash\{0\}$
in $\kd$ (the first $\pi_D$ is not preceeded by a~$\beta$).

$(ii)$ This follows as a global statement from part $(i)$, see also 
{Selmer} \cite[VI,4]{Sel} ({Golomb}'s Theorem).

$(iii)$ Every jump produces exactly one nonzero symbol in $\kd$. 
In the case $m_a(2n)>0$, the nonzero corresponding to the last jump is $\kd(a)_{n+m(2n)/2}$ 
{\it after} $\kd(a)_{i,i=1\dots n}$ and hence must be
accounted for by adding $\delta=1$.

$(iv)$ The statement follows from $(iii)$ and Theorem 14 (see also
\cite[5.3]{Car}).\hfill$\Box$

{\it Theorem $18$}\
({Carter} \cite{Car} for $\ff_2$,
{Niederreiter} \cite{Nr90} for $\ff_q$, $\forall\ q$)\\
{\it
The expectation for the jump complexity is\\
$(i)$ for even $t$
$${\overline {J(t)}} = \frac{t}{2}\cdot\frac{q-1}{q} + \frac{1}{q+1}
- \frac{1}{(q+1)\cdot q^t},$$
$(ii)$ and for odd $t$
$${\overline {J(t)}} = \frac{t}{2}\cdot\frac{q-1}{q} +
\frac{q^2+1}{2q\cdot(q+1)} - \frac{1}{(q+1)\cdot q^t}.$$
}

\bw\
$(i)$ The first $t$ bits of $\kk(a)$ contain at least
$\frac{t}{2}$ bits from $\kd(a)$, of whose on average
$\frac{t}{2}\cdot\frac{q-1}{q}$ are nonzero and thus mark a jump.
Furthermore, if $m_a(t) > 0$ there was another jump in $\kd(a)$
after  $\frac{t}{2}$. Since $t$ is even, we obtain with Corollary 13 that
$$
\sum_{m=2,4,6\dots t} (q-1)\cdot q^{t-m}
= (q-1)\cdot \sum_{m'=0}^{t/2-1} (q^2)^{m'}
= (q-1)\cdot \frac{(q^2)^{t/2} -1}{q^2-1}
= \frac{q^t -1}{q+1}\\
$$
of all $q^t$ sequences in $\ff_q^t$ will deliver one more jump, thus in total
$${\overline {J(t)}} = \frac{t}{2}\cdot\frac{q-1}{q} + \frac{1}{q^t}
\cdot \frac{q^t-1}{q+1}
= \frac{t}{2}\cdot\frac{q-1}{q} + \frac{1}{q+1}
- \frac{1}{(q+1)\cdot q^t}
$$
$(ii)$ Similar to $(i)$ we get $\frac{t+1}{2}\cdot\frac{q-1}{q}$ jumps
on average and one more jump for positive odd $m_a(t)$. With
$$
\sum_{m=3,5,7\dots t} (q-1)\cdot q^{t-m}
= (q-1)\cdot \sum_{m'=0}^{(t-3)/2} (q^2)^{m'}
= (q-1)\cdot \frac{(q^2)^{(t-1)/2} -1}{q^2-1}
= \frac{q^{t-1} -1}{q+1}
$$
we thus obtain
$${\overline {J(t)}} = \frac{t+1}{2}\cdot\frac{q-1}{q} + \frac{1}{q^t}
\cdot \frac{q^{t-1}-1}{q+1}
= \frac{t}{2}\cdot\frac{q-1}{q} 
+\frac{q-1}{2q} 
+\frac{1}{q(q+1)} 
- \frac{1}{(q+1) q^t}\hspace*{1 mm}\Box$$
\\\\\\
{\it F.\ Recurrence Times, $m$--Pattern Frequencies}
\\\\
For $k,l \in\zz$ let
$$\Delta(k,l) =\sum_{\tau=1}^\infty \tau\cdot
p\bigg(m(t+\tau)=l\ |\ m(t)=k\land m(t+1),\dots, m(t+\tau-1)\neq l\bigg)$$
denote the {\it average recurrence time} to go from $m(t)=k$ to $m(t+\tau)=l$
$($by the Translation Theorem 9 the probabilities are independent of $t)$.

{\it Theorem $19$}
\[\begin{array}{rllllll}
{(i)}   &\Delta(k,k)  &=& \frac{1}{q-1}\cdot
2\cdot q^{|k-\frac{1}{2}|+\frac{1}{2}},
&k\in\zz\\
{(ii)}  &\Delta(k,l)  &=& k-l,&k > l \geq 0\\
{(iii)} &\Delta(l,k)  &=& \Delta(k,k) -\Delta(k,l),&  k > l \geq 0\\
{(iv)}  &\Delta(-k,0) &=& k+2\cdot q / (q-1),&k \in \nn_0\\
{(v)}   &\Delta(-k,l) &=& \Delta(-k,0) -l,&0 < l \leq k\\
{(vi)}  &\Delta(k,-k) &=& 2\cdot q\cdot (q^k-1)/(q-1)&k \in\nn\\
{(vii)} &\Delta(k,-l) &=& \Delta(l,-l)+k-l,&k,l \in \nn_0\\
{(viii)}&\Delta(-k,-l)&=& \Delta(-l,-l)+k-l,&0<l\leq k\\
{(ix)}  &\Delta(-k,l) &=& \Delta(0,l) - \Delta(0,-k),&0<k<|l|\\
{(x)}   &p\big(m(t)=k\big) &=& \Delta(k,k)^{-1}\\
\end{array}\]

\bw

$(i)$ To obtain an $m(t)=k>0$, a polynomial of degree $d \geq k$ must occur, 
hence $\kd$ must include a pattern $0^{k-1}\alpha, \alpha\neq 0$ 
(see 16$(ii)$).
This pattern has probability $\frac{q-1}{q^k}$, its recurrence time in $\kd$ 
is thus
$\frac{q^k}{q-1}$. Since an equally large part in  $\kc$ must be passed, we
have $\Delta(k,k) = 2q^k/(q-1)$ for $k\in\nn$.
Negative $m(t)=-k\leq 0$ similarly require the pattern $\alpha0^k$, whence
$\Delta(-k,-k)=2q^{k+1}/(q-1)$ for $k\in\nn_0$.

$(ii)$ From $m(t)=k>0$ the first $k-l$ symbols lead to $m(t+k-l)=l\geq 0$.

$(iii)$ Every path from $k$ to $k$ passes $l$, hence
$\Delta(k,k) = \Delta(k,l) + \Delta(l,k)$.

$(iv)$ We jump after $n$ steps with probability $\frac{q-1}{q^n}$ and
then reach $m=0$ after a total of $k+2n$ steps. Hence
$\Delta(-k,0) = \sum_{n=1}^\infty \frac{q-1}{q^n}\cdot(k+2n) =
k+\frac{2q}{q-1}$.

$(v)$ is a consequence of $(iv)$ and $(ii)$, since
$\Delta(-k,0) = \Delta(-k,l) + \Delta(l,0)$ for $1 \leq l\leq k$.

$(vi)$ We have $\Delta(k,-k) = k + \Delta(0,-k)$ and
$\Delta(-k,-k) = \Delta(-k,0) + \Delta(0,-k)$, hence
$\Delta(k,-k) = k +\Delta(-k,-k) - \Delta(-k,0)
=k+\frac{2q^{k+1}}{q-1}-(k+\frac{2q}{q-1}).$

$(vii)$ For $k\geq l$ we have to add  $k-l$ steps before $\Delta(l,-l)$,
for $k<l$ the first $l-k$ steps have to be omitted.

$(viii)$ Since $|-k|\ \geq \ |-l|$, the path leads through zero, hence
$\Delta(-k,-l) = \Delta(-k,0) + \Delta(0,-l)$. The result now follows from
$(iv)$ and $(vii)$.

$(ix)$ To get from $m=0$ to $m=l\in\zz$, we have to pass the values
$m=-1,-2,\dots,-|l|+1$, hence
$\Delta(0,l) = \Delta(0,-k) + \Delta(-k,l)$ f\"ur $1\leq k < |l|$.

$(x)$ This follows from
$$\hspace*{10 mm}p\big(m(t)=k\big)\ =
\lim_{n\rightarrow \infty}\frac{1}{n}\ \cdot
\ |\{t\ |\ m(t)=k,1\leq k\leq n\}| = \Delta(k,k)^{-1}\hspace*{12 mm}\Box$$

{\it Corollary $20$}\ \ \ \ 
{\it
In the binary case the formulae can be considerably simplified.
Remarkably, over $\ff_2$ the value $\Delta(k,l)$
is integral for all $k,l\in\zz$.

$\begin{array}{rlllll}
{(i)}   &\Delta_2(k,k)  &=& 2^{|k-\frac{1}{2}|+\frac{3}{2}}\  ,&k\in\zz\\
{(ii)}  &\Delta_2(k,l)  &=& k-l,&  k > l \geq 0\\
{(iii)} &\Delta_2(l,k)  &=& 2^{k+1}+k-l,& k > l \geq 0\\
{(iv)}  &\Delta_2(-k,0) &=& k+4,& k \in \nn_0\\
{(v)}   &\Delta_2(-k,l) &=& k -l +4,&0 < l \leq k\\
{(vi)}  &\Delta_2(k,-k) &=& 2^{k+2} -4,&  k \in\nn\\
{(vii)} &\Delta_2(k,-l) &=& 2^{l+2}+k-l-4,&k,l \in \nn_0\\
{(viii)}&\Delta_2(-k,-l)&=& 2^{l+2}+k-l,&0<l\leq k\\
{(ix)}  &\Delta_2(-k,l) &=& \Delta_2(0,l) - \Delta_2(0,-k),
&0<k<|l|\hspace{2 cm}\hfill\Box\\\\
\end{array}$
}

{Niederreiter} defines in \cite[Sect.~5]{Nr90} formulae for
the average frequency of
$L(t) = \frac{t + c_0}{2} \land L(t+1) = \frac{t + c_1}{2}$ in the sequence
$L(i)_{i=1}^n $ for $n \rightarrow \infty$. Using $m$--notation this
corresponds to $m(t) = c_0$ and $m(t+1) = c_1 - 1$. We give a formula for
arbitrary patterns $m(t), m(t+1),\dots,m(t+k-1)$.

{\it Theorem $21$}\ \
{\it
Let an $m$--pattern $(m_0,m_1,\dots,m_k)$ be given with
$m_{i+1}\in\{m_i-1, 1-m_i\}$ and $m_i > 0 \Rightarrow m_{i+1} = m_i-1$.
Let\\
$\#( \downarrow) :=
|\{m_i | m_i \leq 0\ \land\ m_{i+1}= m_i - 1, 0 \leq i \leq k-1\}|$ and\\
$\#( \uparrow) :=
|\{m_i | m_i \leq 0\ \land\ m_{i+1}= 1 - m_i, 0 \leq i \leq k-1\}|$.

The probability of occurrence for the pattern $(m_0,m_1,\dots,m_k)$, that is
$$\lim_{N\rightarrow\infty} \frac{1}{N}\cdot|\{t\ | \ 1\leq t \leq N,
m(t+j) = m_j,0\leq j\leq k\}|,$$
is
$$p(m_0,m_1,\dots,m_k) = \frac{1}{\Delta(m_0,m_0)} \cdot
\frac{(q-1)^{\#( \uparrow)}}
{q^{\#( \downarrow)+\#( \uparrow)}}$$
}

\bw\
The conditions make the pattern feasible (all other patterns have probability
zero).
The pattern occurs, whenever $m_0$ occurs with $p(m_0)$
$= \Delta(m_0,m_0)^{-1}$,
and furthermore, if for $m_i \leq 0$ the linear complexity deviation
jumps $\#( \uparrow)$ times (with probability  $p=\frac{q-1}{q}$)
and is decremented $\#( \downarrow)$ times (with probability 
$p=\frac{1}{q}$).\hfill $\Box$
\\\\\\
\centerline{\sc V. L\'evy Classes and Sharp Asymptotic Bounds for $J$ and $m$}
\\\\
As we have seen, the linear complexity deviation and the jump complexity of a
sequence $a$ can be described immediately in terms of $\kd(a)$.
Furthermore, for each length $t$ and any given sequence $b\in\ff_q^t$ the
probability  $p(\kd(a)_{i,i=1\dots t}=b) = q^{-t}$, hence equidistributed.
We may thus convert theorems about the behaviour of {Bernoulli} 
sequences,
like the Law of the Iterated Logarithm, directly into corollaries about the
asymptotic behaviour of  $m_a(t)$ and $J_a(t)$.

The theorems used here are compiled in  {R\'ev\'esz} \cite{Re1}.
Since {R\'ev\'esz} only treats the case $\ff_2$, we also refrain from
utmost generality and consider only binary sequences, anyway the most important
case from a practical point of view. An exception is the {\it Law of the
Iterated Logarithm for the jump complexity} which we show for $\ff_q^\infty,\ q$
an arbitrary prime power.
\\\\
{\it A.\ L\'evy Classes}

We shall use repeatedly the notation ($\forall_\mu\ a\in\ff_q^\infty \dots)$
or {\it $(\mu$--almost all} $a\in\ff_q^\infty$  $\dots)$ to imply
that the statement $\dots$ is valid on a subset  $A\subset \ff_q^\infty$
of measure $\mu^\infty(A)=1$, hence false at most on a set of measure zero.

The functions $m_a$, $L_a$ $J_a$ and similar are defined
on $\nn$ (we ignore the value at zero). When we vary $a$ on $\ff_q^\infty$,
we obtain functions
on $\ff_q^\infty \times \nn$. Let us first examine the partition into an
$a$--invariant part and the oscillation that depends on $a$.
Given a function $f : \ff_q^\infty \times \nn \rightarrow \rr$, we define
its {\it L\'evy classes}, four classes of functions 
({\it i.e.} classes of real--valued sequences) that describe the
asymptotic behaviour of $f$ (\underline upper and~\underline lower \underline
class, resp.):

\bt{rlll}
$(i)$&$ UUC(f) $&$= \{\alpha \in\rr^\nn \ | \ \forall_\mu \ a,
\exists t_0 \in\nn, \forall t>t_0: $&$f(a,t) < \alpha(t)\}$\\
$(ii)$&$ ULC(f) $&$= \{\alpha \in\rr^\nn \ | \ \forall_\mu \ a,
\forall t_0 \in\nn, \exists t > t_0 : $&$f(a,t) \geq \alpha(t)\}$\\
$(iii)$&$ LUC(f) $&$= \{\alpha \in\rr^\nn \ | \ \forall_\mu \ a,
\forall t_0 \in\nn, \exists t > t_0 : $&$f(a,t) \leq \alpha(t)\}$\\
$(iv)$&$ LLC(f) $&$= \{\alpha \in\rr^\nn \ | \ \forall_\mu \ a,
\exists t_0 \in\nn, \forall t>t_0: $&$f(a,t) > \alpha(t)\}$\\
\et

Thus for all choices 
$\alpha_1\in LLC(f)$, 
$\alpha_2\in LUC(f)$,
$\alpha_3\in ULC(f)$, 
$\alpha_4\in UUC(f)$ and for almost all sequences $a\in A^\infty$, 
we have $\alpha_1< f(a) < \alpha_4$ asymptotically,
but $\mu$--almost all sequences will make $f$ oscillate so much as to 
repeatedly leave the interval $(\alpha_2,\alpha_3)$ of unavoidable oscillation.
\\\\
In the sequel we will use the following typical examples for functions $f$:
maximum length of runs of zeroes (that is jump height, degree of partial
denominators, deviation $m$)  in $\kd$,  as well as deviations
$|J_a(t) -\frac{t}{2}\cdot \frac{q-1}{q}|$. For the case $q=2$ there exist very
precise estimates for the  {L\'evy} classes.

The model used by R\'ev\'esz is a discrete Brownian motion on~$\zz$:
Let $X_i\in\{-1,+1\}$  be  random variables with
$p(X_i=+1)=p(X_i=-1)=\frac{1}{2}$. We start at time $t=0$ at zero.
Given a  sequence $b\in\ff_2^\infty$, let
$$S_b(n) = \sum_{t=1}^n (2\cdot b_t - 1) = -n + 2\cdot \sum_{t=1}^n b_t,
\ n\in\nn_0$$
be the deviation from zero after $n$ moves, where $b_t=0$ corresponds to
$X_t=-1$, and $b_t=1$ to $X_t=+1$.
\\\\
{\it B.\  Jump Complexity}

The jump complexity counts the {\it number} of partial denominators in the 
encoding $\kk(a) = \pi(A_1)|\pi(A_2)|\dots$, which is equivalent to the 
number of nonzeroes in
$\kk_D(a) = \pi_D(A_1)|\pi_D(A_2)|\dots = 0^{d_1-1}lc_1|0^{d_2-1}lc_2|\dots$
(using twice as many symbols in $\kk$ as in $\kd$).
Hence we have the following model for $J$ in terms of $S$:

{\it Theorem $22$}\ \
{\it
For every sequence  $a\in\ff_2^\infty$, $b=\kk(a)$, and every length
$2\cdot t, t \in\nn_0$  we have:
$$
J_a(2\cdot t) = \frac{S_b(t) + t}{2} + \delta,
\ \delta = \Big\{
\begin{array}{ll}
0,&m_a(2\cdot t) \leq 0\\
1,&m_a(2\cdot t) >    0\\
\end{array}
$$
}
\bw \
In Theorem  17$(iii)$ we saw
$J_a(2\cdot t) = \sum_{i=1}^t \kd(a)_i + \ \delta$
(we may just sum up instead of counting nonzeroes since we work over $\ff_2$).
Now, identifying $\kd(a)$ with $(b_k)$, the theorem follows from 
$\frac{S_b(t)+t}{2} = \sum_{k=1}^t b_k$.\hspace*{1 mm}\hfill$\Box$

{\it Theorem $23$}\ \
{\it Law of the Iterated Logarithm for tossing a fair coin}
\nopagebreak
{\it
\[
\ba{lll}
f(t) \in UUC(S_a(t)/\sqrt{t}) &\iff&
\sum_{n=1}^\infty \frac{f(n)}{n} \cdot e^{-\frac{f(n)^2}{2}} < \infty\\
f(t) \in ULC(S_a(t)/\sqrt{t}) &\iff&
\sum_{n=1}^\infty \frac{f(n)}{n} \cdot e^{-\frac{f(n)^2}{2}} = \infty\\
f(t) \in LUC(S_a(t)/\sqrt{t}) &\iff&
-f(t) \in ULC(S_a(t)/\sqrt{t})\\
f(t) \in LLC(S_a(t)/\sqrt{t}) &\iff&
-f(t) \in UUC(S_a(t)/\sqrt{t})\\
\ea
\]

}

\bw
The proof goes back to  {Erd\H{o}s} \cite{Erd}, { Feller} \cite{Fel2},
and {Kolmogoroff} \cite{Kol1}. The theorem is [5.2] of {R\'ev\'esz}
\cite{Re1}. \hfill$\Box$

Some example functions bounding $S_a(t)/\sqrt(t)$ show that we can not avoid
oscillations on the order of the ``iterated logarithm'' $\log\log(t)$.

{\it Example }\ \
For all $\varepsilon>0$ we have:
\[\ba{rcrl}
(2\cdot \log \log(t) &+&(3+\varepsilon)\cdot\log\log\log t)^{1/2}
&\in UUC(S_a(t)/\sqrt{t})\\
(2\cdot \log \log(t) &+&\log\log\log t)^{1/2}
&\in ULC(S_a(t)/\sqrt{t})\\
-(2\cdot \log \log(t) &+&\log\log\log t)^{1/2}
&\in LUC(S_a(t)/\sqrt{t})\\
-(2\cdot \log \log(t) &+&(3+\varepsilon)\cdot\log\log\log t)^{1/2}
&\in LLC(S_a(t)/\sqrt{t})\\
\ea\]

From Theorem 23 we now infer a Law of the Iterated Logarithm for the Jump
Complexity in the binary case:\

{\it Theorem $24$}  \ \
{\it
For $\mu$--almost all $a\in\ff_2^\infty$ with $b = \kk(a)$ the jump
complexity $J_a(2t)$ observes:
\[
\ba{lcll}
f(t) \in UUC(S_b(t)/\sqrt{t}) &\Rightarrow&
(\sqrt{t}\cdot f(t)+t)/2 +1 &\in UUC(J_a(2t))\\
f(t) \in ULC(S_b(t)/\sqrt{t}) &\Rightarrow&
(\sqrt{t}\cdot f(t)+t)/2  &\in ULC(J_a(2t))\\
f(t) \in LUC(S_b(t)/\sqrt{t}) &\Rightarrow&
(\sqrt{t}\cdot f(t)+t)/2 +1 &\in LUC(J_a(2t))\\
f(t) \in LLC(S_b(t)/\sqrt{t}) &\Rightarrow&
(\sqrt{t}\cdot f(t)+t)/2  &\in LLC(J_a(2t))\\
\ea
\]

In particular, the classes contain the following functions:
\[\ba{lrccrl}
t/4+1&+\ \sqrt{t}&\cdot&\sqrt{\log \log(t)/4 +
(3/8+\varepsilon)\cdot\log\log\log t}&\in UUC(J_a(t))\\
t/4  &+\ \sqrt{t}&\cdot&\sqrt{\log \log(t)/4 +
1/8\cdot\log\log\log t}&\in ULC(J_a(t))\\
t/4+1&-\ \sqrt{t}&\cdot&\sqrt{\log \log(t)/4 +
1/8\cdot\log\log\log t}&\in LUC(J_a(t))\\
t/4  &-\ \sqrt{t}&\cdot&\sqrt{\log \log(t)/4 +
(3/8+\varepsilon)\cdot\log\log\log t}&\in LLC(J_a(t))\\
\ea\]
}

\bw
By Theorem 22, we can replace $J_a$ by $\frac{S_b(t) + t}{2} + \delta$
and thus obtain with Theorem~23 and the above example, resp.~the statements. 
See also the next theorem for arbitrary finite fields $\ff_q.$ \hfill$\Box$

{\it Theorem $25$}\ \
{\it The Law of the Iterated Logarithm for the Jump Complexity over arbitrary
finite fields  $\ff_q$
}
\[\ba{rcl}
\hspace{-1 cm}{(i)}&\overline{\lim}_{n\rightarrow \infty}\ 
(J_{a}(n)- \frac{n}{2}\cdot\frac{q-1}{q})\ /\ \sqrt{\frac{q-1}{q^2} n\cdot\log\log n} = +1
&\mu-{\rm a.e.}\\
\hspace{-1 cm}{(ii)}&\underline{\lim}_{n\rightarrow \infty}\ 
(J_{a}(n)- \frac{n}{2}\cdot\frac{q-1}{q})\ /\ \sqrt{\frac{q-1}{q^2} n\cdot\log\log n} = -1
&\mu-{\rm a.e.}\\
\ea\]

\bw\
Theorem 22~can be immediately generalized to all prime powers $q$.
Thus we can apply the Law of the Iterated Logarithm for {Bernoulli}
sequences with  $p = \frac{1}{q}$ as proportion of zeroes
(we do not descriminate between nonzero symbols)
(see {Feller} \cite{Fel1}) to obtain  (i) and (ii).\hfill$\Box$
\\\\
{\it C.\ Linear Complexity Deviation $m$ }

We now consider the linear complexity deviation $m$.
$m$ takes on its maximum values at a jump (except for rational $a$,
but we can ignore the set $\kk^{-1}(\ser\cap\ff_q[x])$  of measure zero).
Hence the length of zero runs in $\kd(a)$ is of importance, since every
$\pi_D$ consists of a zero run (of length $d-1\in \nn_0$)  and the leading
coefficient that makes $m$ jump. We first define
the length of the largest uninterrupted sequence of  $-1$
in $(X_i)_{i=1,\dots,n}$ or of zeroes in $b_1,\dots,b_n$, resp.~as
$$Z_b(n) := \max_{0\leq r\leq n} \Big\{r=\max_{0\leq k\leq n-r}
\Big(S_b(k) - S_b(k+r)\Big)\Big\}$$

{\it Theorem $26$}\ \
{\it
Let  $a\in\ff_2^\infty$ with  $b = \kd(a)$ and  $m_a(2n)=0$,
then :
$$\max_{t\leq 2n}\ |m_a(t)|\ = Z_b(n)+1$$
}

\bw\ \ 
The largest value $|m_a(t)|$ occurs after the longest zero run in $\kd$,
which is of length $\max_{t\leq 2n}\ |m_a(t)| - 1$ and terminates with 
a leading coefficient, since $m_a(2n)=0$. $Z_b$ just gives that longest 
run length, and at $t=2n$, $\kd$ contains $n$ symbols. \hfill$\Box$

{\it Theorem $27$}\ \ \   {\it L\'evy classes for $Z(n)$}

$$\sum_{n=1}^\infty 2^{-f(n)} < \infty \iff f(n) \in UUC(Z(n))$$
$$\sum_{n=1}^\infty 2^{-f(n)} = \infty \iff f(n) \in ULC(Z(n))$$
$$f(n) = \lfloor \log_2(n) -\log_2\log_2\log_2(n)+\log_2\log_2(e) -1
+\varepsilon\rfloor\in LUC(Z(n)),\forall \varepsilon>0$$
$$f(n) = \lfloor \log_2(n) -\log_2\log_2\log_2(n)+\log_2\log_2(e) -2
-\varepsilon\rfloor\in LLC(Z(n)),\forall \varepsilon>0$$

\bw\
See  {Erd\H{o}s} and {R\'ev\'esz} \cite{ER}, 
{R\'ev\'esz} \cite{Re2}.\hfill$\Box$

{\it Theorem $28$}\ \ \  {\it L\'evy classes for 
$m^+(n) := \max_{t\leq n}|m(t)|$}
$$\sum_{n=1}^\infty 2^{-f(n)} < \infty \iff f(n)
\in UUC(m^+(n))$$
$$\sum_{n=1}^\infty 2^{-f(n)} = \infty \iff f(n)
\in ULC(m^+(n))$$
$$f(n) = 1+\lfloor \log_2(\frac{n}{2}) -\log_2\log_2\log_2(
\frac{n}{2})+\log_2\log_2(e) +\varepsilon\rfloor$$
{\hspace*{3 mm}\hfill{$\in LUC(m^+(n)),\ \forall \varepsilon>0$}}
$$f(n) = 1+\lfloor \log_2(\frac{n}{2}) -\log_2\log_2\log_2(
\frac{n}{2})+\log_2\log_2(e)-2-\varepsilon\rfloor$$
{\hspace*{3 mm}\hfill{$\in LLC(m^+(n)),\ \forall \varepsilon>0$}}

\bw
The theorem follows from Theorems 26 and 27, where the convergence of the sums
does not depend on the constant  $+1$ or the fact that the function $f(n/2)$
is sampled twice as often (both supply only a factor $2^{\pm 1}$).
The statements about $UUC$ and  $ULC$  already appear as 
\cite[Th.~8, 9]{Nr88}.\hfill$\Box$

Theorem 28 allows us to show that a fixed bound on $m$  generally is too 
restrictive, but logarithmic growth is feasible for $m$. We therefore define
the notions of {\it perfect} and {\it good} profiles.\\
A sequence $a\in\ff_q^\infty$ is called {\it $d$--perfect} for a $d\in\nn$ if 
$\forall t\in\nn:\ |m_a(t)|\leq d$.\\
A sequence  $a$ has a  {\it good  linear complexity profile}, if
$$\exists \ C\in\rr, \forall n\in\nn:\ |m_a(n)| \leq 1 + C\cdot \log n.$$

{\it Corollary $29$} \ \\ 
{\it
$(i)$ There are  $\mu$--almost no sequences with $d$--perfect  linear
complexity profile.\\
$(ii)$ $\mu$--almost all sequences  have a good linear complexity profile
}

\bw

$(i)$ For all $d\in\nn$, $f(n)=d$ becomes smaller than every sequence in $LLC$,
which itself is superated by $\mu$--almost all sequences from some $n_0$ on.

$(ii)$ For $f(n) := 2\cdot \log_2(n)$ the sum $\sum_{n=1}^\infty 2^{-f(n)}$ 
converges, and hence for $\mu$--almost all sequences
$a$ we have $m_a(t)<f(t)\in UUC$ for all $t>t_0(a)$ for some
$t_0(a)$.\hfill$\Box$

{\it Theorem  $30$}\  \cite[Th.~2]{Nr90a}\ \
{\it
Let $f,g$ be functions on $\nn$ with  $\lim_{n\to\infty} f(n)=
\lim_{n\to\infty} g(n)=\infty$. Then:
$$\forall_\mu\ \ a\in\ff_q^\infty\ : \ -f(n)\leq m_a(n)\leq g(n)$$
}
\bw
See {Niederreiter} \cite[Th.~2]{Nr90a}.\hfill$\Box$

{\it Remark }\
Theorems 27 and 30 together show:\ \
The largest linear complexity deviation ever occurring
is of the order  $\log_2 n$ for almost all sequences ---
however, for almost all sequences from a fixed $n_0$
onwards the linear complexity deviation is only of
the order $f(n):=\log_2^{(k)} n$ for arbitrarily large~$k$.

{\it Theorem $31$}

{\it
$(i)$ For the lengths $Z_2(n), Z_3(n), \dots$ of the second,
third $\dots$ largest run
of zeroes {Deheuvels} {\rm \cite{Deh}} has found the following functions
$($where $\log_2^{(j)}:={\underline{\log}}_2(log_2^{(j-1)})$ with
${\underline{\log_2}}(x) = 0 $ for $x<1$,
${\underline{\log_2}}(x) = \log_2(x) $ for $x\geq 1$.

 For all $k\in\nn, r \geq 2, \varepsilon>0$ we have:
\begin{eqnarray*}
f(n) &=& \log_2(n) +\frac{1}{k}\cdot (\log_2^{(2)}(n)+\dots+\log_2^{(r-1)}(n)
+ (1+\varepsilon)\log_2^{(r)}(n))\\
&&   \in UUC(Z_k(n))\\
f(n) &=& \log_2(n) +\frac{1}{k}\cdot (\log_2^{(2)}(n)+\dots+\log_2^{(r-1)}(n)
+ \log_2^{(r)}(n))   \in ULC(Z_k(n))\\
f(n) &=& \lfloor \log_2(n) -\log_2\log_2\log_2(n)+\log_2\log_2(e)
+\varepsilon\rfloor   \in LUC(Z_k(n))\\
f(n) &=& \lfloor \log_2(n) -\log_2\log_2\log_2(n)+\log_2\log_2(e)-2
-\varepsilon\rfloor  \in LLC(Z_k(n))\\
\end{eqnarray*}
$(ii)$ Hence for the second etc.~largest degree $d_k(t)$ in the continued
fraction expansion, we obtain:
}
\begin{eqnarray*}
f(t) &=&1+ \log_2(\frac{t}{2}) +\frac{1}{k}\cdot (\log_2^{(2)}(\frac{t}{2})+
\dots+\log_2^{(r-1)}(\frac{t}{2}) + (1+\varepsilon)\log_2^{(r)}(\frac{t}{2}))\\
&&\in UUC(d_k(t))\\
f(t) &=&1+ \log_2(\frac{t}{2}) +\frac{1}{k}\cdot (\log_2^{(2)}(\frac{t}{2})+
\dots+\log_2^{(r-1)}(\frac{t}{2})
+\log_2^{(r)}(\frac{t}{2}))   \in ULC(d_k(t))\\
f(t) &=&1+ \lfloor \log_2(\frac{t}{2}) -\log_2\log_2\log_2(\frac{t}{2})+
\log_2\log_2(e)
+\varepsilon\rfloor   \in LUC(d_k(t))\\
f(t) &=&1+ \lfloor \log_2(\frac{t}{2}) -\log_2\log_2\log_2(\frac{t}{2})+
\log_2\log_2(e)-2
-\varepsilon\rfloor  \in LLC(d_k(t))\\
\end{eqnarray*}
\bw\
$(i)$ See {Deheuvels} \cite{Deh} and {R\'ev\'esz} \cite[S.~61]{Re1}.
Observe that  $LUC, LLC$  do not depend on $k$.\\
$(ii)$ follows from $(i)$ with Theorem 26, compare the proof to 
Theorem~28.~\hfill~$\Box$
\\\\\\
\centerline{\sc VI. 2--adic Span and Complexity}
\\\\
Klapper and Goresky \cite{KG} \cite{KG1} introduced another measure to assess the
(non--)\-randomness of a bit string, the representation of 
$(a_1,a_2,a_3,\dots)$
as a 2--adic integer $a=\sum_{i=1}^\infty a_i2^{i-1}\in\zz_2$ and 
its approximations by rational numbers from $\qq$,
as given in Mahler \cite{Mah} and De Weger \cite{deW}.

Klapper and Goresky view a sequence $(a_1,a_2,\dots)\in\f2o$ as a more and
more precise description of a $2$--adic integer in $\zz_2$ (the base field
now must be prime, not just a prime power, and we shall treat only  $p=2$).
For every finite prefix $(a_1,\dots,a_k)$, we obtain the number
$a^{(k)} := \sum_{i=1}^k a_i2^{i-1} \in \zz\subset \zz_2$.
We are now interested in describing the number $\ak$  by a fraction
$\frac{p_k}{q_k}$ with the condition
$q_k\cdot \ak \equiv p_k\ \mod\ 2^k$.

We need some definitions:

For $k\in\nn, a_1,\dots, a_k \in\ff_2,$ let 
$\ak = \sum_{i=1}^k a_i2^{i-1} \in \zz$. We define the lattice 
(or $\zz$--module) 
${\cal L}_a(k) = \{(p_k,q_k)\in\zz^2\ | 
\ q_k\cdot \ak \equiv p_k \ \mod\  2^k\}$
and set ${\cal L}_a'(k) = {\cal L}_a(k) \backslash \{(0,0)\}$.

For $(p,q) \in \zz^2$, let $\Phi(p,q) = \max(|p|,|q|)$.

Let $(c_k,d_k)$ be a minimal pair from ${\cal L}_a(k)$ in the sense of
$\Phi((c_k,d_k)) \leq \Phi((p_k,q_k))$ for all $(p_k,q_k)\in {\cal L}_a'(k)$.

We call the sequence $(c_k,d_k)$ the minimal approximating sequence of $a$.

The 2--adic complexity now is defined as
$\phi_2(a,k) := \log_2(\Phi(c_k,d_k))\in\rr$ in \cite{KG}.
However, we are interested here in an isometric model of 2--adic
approximation. Thus we define
${\bf A }\colon \f2o \to \f2o$ as\\
$$\ka(a_1,a_2,\dots)_k = \left\{
\ba{ll}
0,&\mbox{\rm if\ }(c_k,d_k) = (c_{k-1},d_{k-1})\\
1,&\mbox{\rm if\ }(c_k,d_k) \neq (c_{k-1},d_{k-1})\\
\ea
\right.
$$
for $k\in\nn$, where we set $(c_0,d_0) := (0,1)$.

{\it Theorem $32$}\ \
{\it
$\ka$ is an isometry on $\f2o$.
}

\bw\
Let $a,b\in\f2o$ with
$a_i=b_i, 1\leq i <k$ and $a_k = 1-b_k$.
Then $\ka(a)_i = \ka(b)_i$ for $1\leq i< k$, since both sequences lead to 
the same sequence $(c_i,d_i)$ of rational approximations.

However, from
$d_{k-1}\cdot \sum_{i=1}^{k-1} a_i2^{i-1} \equiv c_{k-1}\ \mod\ 2^{k-1},$
we may infer\\
$d_{k-1}\cdot \sum_{i=1}^{k-1} a_i2^{i-1} + \delta\cdot 2^{k-1} \equiv c_{k-1}\ 
\mod\ 2^{k},$ for a $ \delta\in\{0,1\}$ and thus\\
$d_{k-1}\cdot \sum_{i=1}^{k-1} a_i2^{i-1} + (1-\delta)\cdot 2^{k-1} \not\equiv c_{k-1}\ 
\mod\ 2^{k}$.

Thus
$\ka(a_1,\dots,a_{k-1},\delta,\dots)_k = 0$ and
$\ka(a_1,\dots,a_{k-1},1- \delta,\dots)_k = 1$.

One of these corresponds to $\ka(a)_k$, the other to $\ka(b)_k$, hence
$|{\de}\ka(a)-\ka(b)|{\de} = -k = |{\de}a-b|{\de}$ and $\ka$ is an isometry.\hfill $\Box$

We further define the 2--adic jump complexity $J_A$ to count the number of changes 
in the 2--adic complexity profil:
$J_A(a)(n) := \sum_{i=1}^n \ka(a)_i$,
which should behave like $J_A(n)\approx \frac{n}{2}$. 
In analogy to $m$, we thus define the 
2--adic jump complexity deviation $m_A(n) := 2\cdot J_A(n) - n\in\zz$.

{\it Theorem $33$}\ \
{\it
$m_A(t)$ and $S_b(t)$ have the same average and asymptotic behaviour, 
precisely
$$f\in LLC\dots UUC(m_A(t))\Longleftrightarrow 
f\in LLC\dots UUC(S_b(t))$$
}

\bw\ 
$J_A(t)$ and the sum $\sum_{i=1}^t b_i$ of the coin tossing sequence $b_t$ 
both represent the same behaviour, summing up a $p=\frac{1}{2}$ Bernoulli 
experiment.
With $m_A(n) = 2\cdot J_A(n)-n$ and $S_B(n) = 2\cdot\sum_{i=1}^n b_i -n$ 
the result follows.~\hfill~$\Box$

We have thus 

{\it Corollary $34$}\ \ 
{\it
Law of the Iterated Logarithm for the $2$--adic Jump Complexity Deviation

\[\ba{rcl}
\hspace{-1 cm}{(i)}&\overline{\lim}_{n\rightarrow \infty}\ 
m_A(n)\ /\ (\sqrt{2n\cdot\log\log n}) = +1
&\mu-{\rm a.e.}\\
\hspace{-1 cm}{(ii)}&\underline{\lim}_{n\rightarrow \infty}\ 
m_A(n)\ /\ (\sqrt{2n\cdot\log\log n}) = -1
&\mu-{\rm a.e.}\\
\ea\]
}

{\it Remark} \ \
Whereas the original 2--adic complexity $\phi$ is not usually integral,
$J_A$ and $m_A$  put us again in the world of coin tossing. 
However, in order to compute
\ka, until now we have to iteratedly calculate $\Phi$ and derive $\ka$
from the sequence of convergents.

{\it Open Problem} \ \

Can we calculate \ka$(a)_{i,i=1\dots n}$ with bit complexity $O(n^2)$ ?

Can we even calculate the shifted $2$--adic profiles  
\ka$(a_1\dots a_n)$, \ka$(a_2\dots a_n)$
$\dots$  \ka$(a_{n-1}\dots a_n)$ together in $O(n^2)$ time ?
\\\\\\
\centerline{\sc VII. Conclusion}
\\\\
We have shown that both the expansion of formal power series into their
continued fraction expansion and the approximation of 2--adic numbers by
rationals induce an isometry on $\f2o$, $\kk$ and $\ka$, resp.

We modelled linear and jump complexity as well as 2--adic jump complexity
via Bernoulli experiments, and applied the known general bounds to this
model to derive sharp bounds (L\'evy classes) for $J, m,$ and $J_A$.

We gave an adaptation of the Berlekamp--Massey--algorithm that implements the
continued fraction expansion exactly as for the reals to obtain~\kk.
\\\\\\
\centerline{\sc Acknowledgement}

I want to heartfully thank Harald Niederreiter of the Austrian 
Academy of Sciences and National University of Singapore for introducing me 
to the subject and many stimulating and fruitful discussions, since 
accepting me as a PhD student a decade ago.

\end{document}